\documentclass[doublecol]{epl2}
\usepackage{amsmath}
\usepackage{amssymb}
\usepackage{mathtools}
\usepackage{txfonts}
\UseRawInputEncoding

\title{Role of the Casimir force in micro- and nanoelectromechanical
pressure sensors}
\shorttitle{Role of the Casimir force in micro- and nanoelectromechanical
pressure sensors}

\author{G.~L.~Klimchitskaya\inst{1,2} \and A.~S.~Korotkov\inst{2}
\and V.~V.~Loboda\inst{2} \and V.~M.~Mostepanenko\inst{1,2,3}
 }
\shortauthor{G.~L.~Klimchitskaya \etal}

\institute{
  \inst{1}Central Astronomical Observatory at Pulkovo of the
Russian Academy of Sciences, Saint Petersburg,
196140, Russia\\
\inst{2}Peter the Great Saint Petersburg
Polytechnic University, Saint Petersburg, 195251, Russia\\
\inst{3}Kazan Federal University, Kazan, 420008, Russia
}

\abstract{ The Casimir force caused by the electromagnetic fluctuations
is computed in the configurations of micro- and nanoelectromechanical pressure
sensors using Si membranes and either Si or Au-coated Si substrates. It is
shown that if, under the influence of external pressure, the membrane-substrate
separation drops to below 100 nm, the Casimir force makes a profound effect on
the sensor functioning. There exists the maximum value of external pressure
depending on the sensor parameters such that it finds itself in a state of
unstable equilibrium. For this and larger pressures, the Casimir force leads
to a collapse of the sensor, which loses its functionality. For any smaller
external pressures, there exist two equilibrium positions, one of which is
unstable and another one is stable, at smaller and larger membrane-substrate
separations, respectively. The latter can be safely used for the pressure
measurements. Possible applications of the {obtained}
results in the design of
micro- and nanoelectromechanical pressure sensors of next generations with further decreased
dimensions are discussed.}

\begin{document} \maketitle

\newcommand{\kb}{{k_{\bot}}}
\newcommand{\xl}{{i\xi_l}}
\newcommand{\ve}{{\varepsilon}}
\newcommand{\okt}{{(\omega,k_t)}}

\section{Introduction}

During the last decades, micro- and nanoelectromechanical system
sensors (MEMS and NEMS) find increased applications for measuring various
mechanical, electromagnetic and optical quantities including velocity,
acceleration, pressure, electric and magnetic fields, light intensity and many
others (see, for instance, the monographs \cite{1,2}). Among these microdevices,
an important place is occupied by different modifications of the pressure
sensors using mechanical, capacitive, piezoelectric, optical and other techniques
(see, e.g., recent articles \cite{2.1,2.2,2.3,2.4,2.5,2.6,2.7,2.8,2.9,2.10} and
reviews  \cite{3,4,5,6,7,8,9}).

The characteristic sizes of micro- and nanoelectromechanical sensors and their elements
tend to decrease from hundreds of micrometers to micrometers and then to
hundreds and even tens of nanometers, where the effects of fundamental physics
come into play. Although {nowadays} the major role in the sensor operation is
played by the electric force, with shrinking sensor dimensions to below a
micrometer the Casimir force \cite{10} induced by the electromagnetic
fluctuations becomes important. This force acts between any uncharged elements
of sensors and far exceeds the characteristic electric forces at separations
below one hundred of nanometers.

In relation to micro- and nanosensors, the Casimir force can play both a
harmful and beneficial role. Thus, microsensors often lose their functionality
when their moving parts stick to each other under an impact of the Casimir
force \cite{11,12,12a,12b}. On the other hand, it was shown \cite{13,14} that the
Casimir force can replace the electric force as a driving force that ensures
the functioning of a microdevice.

Recent progress in precise measurements of the Casimir force between metallic
\cite{14.1,14.2,14.3,14.4,14.5,14.6,14.7,14.8,14.9,14.10,14.11,14.12,14.13}
and semiconductor \cite{14.14,14.15,14.16,14.17,14.18,14.19,14.20} test bodies
(see also the reviews \cite{15,16,17,18,19}) increased interest in studying the role of
Casimir forces in various microdevices. Thus, the stability of microdevices
actuated by the Casimir force was investigated in \cite{20,21}. The impact
of surface roughness and phase transformations in microdevices controlled by
the Casimir force was considered in \cite{22,23,24,25,26}. The silicon
micromechanical chip driven by the Casimir force was created in \cite{27,28}.
The optical switching of a graphene mechanical switch and the optical
chopper, both driven by the Casimir force, were suggested in \cite{29,30},
respectively. The commercial Casimir-driven microelectromechanical sensors,
capable of monitoring of biomagnetic fields, were elaborated in \cite{31,32}.
In parallel with these studies, an impact of the Casimir force on the pull-in
instability of microdevices using molecular dynamics simulations was
considered \cite{33}.

In this Letter, we investigate the role of Casimir force in pressure
microsensors, which was not considered in the literature so far. This is
done using the model example of simple pressure sensor with a plane membrane
whose weight is balanced by a spring suspension system characterized by some
effective spring constant. Under an impact of external pressure, the membrane
approaches the underlying substrate for a sufficiently small distance where
the Casimir force comes into play.

Below we consider two configurations of the pressure microsensors. In the
initial position of the first one, the weight of the membrane is balanced
by the elastic force arising from some extension of supporting springs.
Then, the external pressure is applied to the membrane. An additional
extension of the spring system, which is measured either mechanically or
optically, is determined by the combined action of the external and
Casimir pressures. In the second configuration, an initial position of
the membrane is the same, but an additional extension of the spring system
occurs under the impact of external pressure, Casimir pressure, and also
electric pressure arising due to the applied potential difference. This
allows determination of the membrane displacements by means of
capacitive measurements.

It is shown that in both configurations there exists the maximum value
of the measured pressure such that the sensor membrane reaches the
separation of an unstable equilibrium above a substrate and finally sticks
it. This value depends on the spring constant and the initial height of a
membrane above a substrate. According to the results obtained, for smaller
values of the measured pressure there are two equilibrium positions of
the sensor membrane above a substrate, one of which is unstable and another
one is stable. When obtaining these results, computations of the Casimir
force were performed by means of the Lifshitz theory for the cases of a Si
membrane and a Si substrate and a Si membrane and an Au-coated Si substrate
using the tabulated optical data for the complex refraction indices of
Si and Au. Applications of the obtained results in new generations of
pressure micro- and nanoelectromechanical sensors with further reduced
dimensions {are discussed}.

\section{Casimir force in configurations of MEMS and NEMS pressure sensors}

\begin{figure}
\onefigure{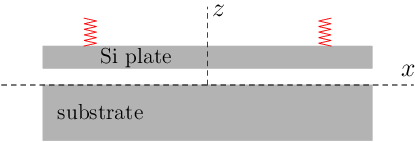}
\caption{Configuration of a Si membrane suspended by the spring system above a thick
plate.}
\label{fg1}
\end{figure}

We consider the Si membrane of $L=1000\,\,\muup$m length,  $D=200\,\,\muup$m
width, and $H=30\,\,\muup$m thickness spaced at some height $h$ above an
entirely Si or an Au-coated Si substrate (see fig.~\ref{fg1}). The area of the
membrane $S=LD=2\times 10^5\,\,\muup\mbox{m}^2$.
Let the position of a membrane is fixed by some spring system with an effective
spring constant $k$. We assume that in the absence of external pressure the
equilibrium height, where the weight of the membrane is counter balanced by the
elastic force, is $h=20\,\,\muup$m. At so large separation, the Casimir force
$P_C(z)$ between a membrane and a substrate is negligibly small. However,
in the presence of external pressure $P$ acting on the upper side of the
membrane, it approaches the substrate. As a result, the role of the Casimir
force increases. This is the first configuration considered below.

The second configuration of MEMS and NEMS pressure sensors is similar to the first
one but, in addition to the elastic and Casimir forces, there is an electric force created
by the potential difference $V_0$ applied between a membrane and a substrate.
The often used material for manufacturing MEMS and NEMS pressure sensors is the
phosphorus-doped Si. An Au coating on the substrate of about $100\,\,\muup$m
thickness helps to increase its conductivity.

The Casimir force in both configurations can be computed by means of the Lifshitz
theory \cite{15,17,34,34a,34b}. Taking into account that the membrane area is much larger
than the separation to the substrate squared, it can be considered as infinitely large.
The thicknesses of both a membrane and a substrate (and also of an Au coating) are
large enough in order they can be considered as infinitely thick \cite{17}.

Under these conditions, the magnitude of the Casimir pressure between a membrane and
a substrate is given by the Lifshitz formula \cite{15,17}
\begin{eqnarray}
&&
P_C(z)=\frac{k_BT}{\pi}\sum_{l=0}^{\infty}{\vphantom{\sum}}^{\prime}\!\!
\int_{0}^{\infty}\!\!\!q_l\kb d\kb
\nonumber \\
&&~~~~~~~~
\times
\left[
\frac{r_{\rm TM}^{(1)}(\xl,\kb)r_{\rm TM}^{(2)}(\xl,\kb)\,e^{-2zq_l}}{1-
r_{\rm TM}^{(1)}(\xl,\kb)r_{\rm TM}^{(2)}(\xl,\kb)\,e^{-2zq_l}} \right.
\nonumber \\
&&~~~~~~~~~~~~
\left.+\frac{r_{\rm TE}^{(1)}(\xl,\kb)r_{\rm TE}^{(2)}(\xl,\kb)\,e^{-2zq_l}}{1-
r_{\rm TE}^{(1)}(\xl,\kb)r_{\rm TE}^{(2)}(\xl,\kb)\,e^{-2zq_l}} \right],
\label{eq1}
\end{eqnarray}
\noindent
where
$k_B$ is the Boltzmann constant, $T$ is the temperature,
$q_l^2=k_{\bot}^2+\xi_l^2/c^2$,  $\kb$ is the magnitude of the wave vector
projection on the membrane plane, $\xi_l=2\pi k_BTl/\hbar$ with $l=0,\,1,\,2,\,\ldots$
are the Matsubara frequencies, and prime on the summation {sign}
divides by 2 the term
with $l=0$. The reflection coefficients on a membrane and on a substrate for the
transverse magnetic (TM) and transverse electric (TE) polarizations of the
electromagnetic field are given by
\begin{eqnarray}
&&
r_{\rm TM}^{(1,2)}(\xl,\kb)=
\frac{\ve_l^{(1,2)}q_l-k_l^{(1,2)}}{\ve_l^{(1,2)}q_l+k_l^{(1,2)}},
\nonumber \\
&&
r_{\rm TE}^{(1,2)}(\xl,\kb)=
\frac{q_l-k_l^{(1,2)}}{q_l+k_l^{(1,2)}},
\label{eq2}
\end{eqnarray}
\noindent
where $\ve_l^{(1,2)}=\ve^{(1,2)}(i\xi_l)$ are the dielectric permittivities of
membrane and plate materials computed at the Matsubara frequencies and
\begin{equation}
k_l^{(1,2)}=\left(k_{\bot}^2+\ve_l^{(1,2)}\frac{\xi_l^2}{c^2}\right)^{1/2}.
\label{eq3}
\end{equation}
\noindent
The values of $\ve_l^{(1,2)}$ are found by means of the Kramers-Kronig relations
using the tabulated optical data for Si and Au \cite{35}.
The latter were extrapolated down to zero frequency by means of the plasma model.
It was shown \cite{15,17,19} that this extrapolation agrees with all precise
measurements of the Casimir force in spite of the fact that the plasma model
does not take into account the relaxation of conduction electrons.
According to the recent results \cite{36}, approximately the same values of the
Casimir force in the limits of measurement errors are obtained if the extrapolation
of the optical data in the region of propagating waves and TM evanescent waves
is made by the dissipative Drude model. It was concluded that in the region of
TE evanescent waves the Drude model describes the response of metals to the
low-frequency electromagnetic field incorrectly \cite{36}. Keeping in mind
that in this region the Drude model lacks of experimental confirmation,
the new independent test was proposed \cite{37,38} aiming to confirm this
conclusion.

\section{Stability of MEMS and NEMS pressure sensors}

First we consider the pressure microsensor using a Si membrane and a Si substrate
with the dimensions indicated above. Let the spring constant $k$ be determined
from the condition $kh=P_0S$ where $P_0=35~$kPa, i.e., $k=350~$N/m.
Note that the pressure region from 5~kPa to 40~kPa is characteristic for the
low pressure measurements in liquids, for example, blood pressure.
A pressure of 35~kPa corresponds to a blood pressure of 263~mmHg  \cite{liq1,liq2}.
With this $k$, in the absence of the Casimir
force, the external pressure $P_0$ brings the membrane into a contact with the
substrate.

In the presence of the Casimir pressure, the equilibrium position of a membrane
at the height $z<h$ above a substrate can be determined from the equation
\begin{equation}
f(z)\equiv\frac{kh}{S}-P-\frac{kz}{S}=P_C(z),
\label{eq4}
\end{equation}
\noindent
where $P$ is the magnitude of the applied external pressure.

In Table~1, column 2 we present several typical values of the Casimir pressure
computed by eqs.~(\ref{eq1})--(\ref{eq3}) at $T=300~$K with
$\ve_l^{(1)}=\ve_l^{(2)}=\ve_l^{\rm Si}$. Computations were
performed using the optical data of high resistivity (dielectric) Si \cite{35}.
It was shown \cite{14.16,14.17,14.19,14.20,17} that the presence of free charge carriers,
whose concentration is below the critical value at which the phase transition of Si to the
metallic state occurs, has no effect on the Casimir force at short separations considered here.

\begin{figure}[h]
\onefigure{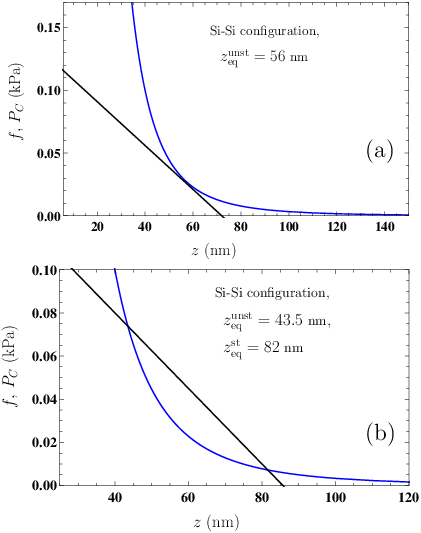}
\caption{Equilibrium positions of the sensor membrane made of Si above a Si substrate
with applied external pressures (a) $P=34.874~$kPa and  (b) 34.85~kPa.
See the text for further discussion. }
\label{fg2}
\end{figure}

In fig.~\ref{fg2}(a,b), the Casimir pressure
between Si surfaces is shown as the function of $z$ by the blue curved line.
In fig.~\ref{fg2}(a), the function $f(z)$ is plotted as the black line for
$P=34.874~$kPa and in fig.~\ref{fg2}(b) --- for $P=34.85~$kPa.
As is seen in fig.~\ref{fg2}(a,b), for the first value of pressure
eq.~(\ref{eq4}) has only one solution, which corresponds to the membrane position
of an unstable equilibrium, $z_{\rm eq}^{\rm unst} =56~$nm, whereas for the
second value this equation has two solutions. One of them,
$z_{\rm eq}^{\rm st} =82~$nm, corresponds to the stable position of
a membrane and another one, $z_{\rm eq}^{\rm unst} =43.5~$nm,
describes the position of an unstable equilibrium.

To summarize, the MEMS and NEMS pressure sensors with the above parameters can be
used for measuring any pressure values below 34.874 kPa, but for this and larger
pressures the sensor membrane will stick to the substrate making the sensor unusable.

Now we consider the case when the Si substrate is coated with an Au layer of
$100~\muup$m thickness. In  this case the Casimir pressure is calculated by
eqs.~(\ref{eq1})--(\ref{eq3}) at $T=300~$K with $\ve_l^{(1)}=\ve_l^{\rm Si}$ and
$\ve_l^{(2)}=\ve_l^{\rm Au}$. The obtained typical values of the Casimir
pressure are presented in Table~1, column 3.

In the air environment, for micro-pressure measurements in the region of 0--5 kPa
\cite{gas1,gas2}, the measured pressures are, typically, much smaller than in liquids.
For example, $P_0 = 3~$kPa (163~dB) is the maximum sound pressure on the tympanic membrane in the human ear called eardrum. For this pressure,
 we arrive at the spring constant  $k=P_0S/h=30~$N/m. With this spring
constant, let us determine the region of stable pressure measurements for the
sensor with an Au-coated substrate and compare it with that for an entirely
Si one.
\begin{table}
\caption{The values of the Casimir pressure between a Si membrane and either
Si or Au-coated Si substrate are shown in columns 2 and 3, respectively, as the
function of separation.}
\label{tab1}
\begin{center}
\begin{tabular}{cll}
$~~z$~(nm)\hspace*{2mm}&\multicolumn{2}{c}{$P_C$~(kPa)}\\
& ~Si-Si \hspace*{5mm}& ~Si-Au \\
~10& 11.3& 11.4\\
~20& 1.03 & 1.22 \\
~30& 0.276 & 0.328\\
~40& 0.0999& 0.120\\
~50 & 0.0430& 0.0539\\
~60&0.0228 & 0.0279\\
~70 & 0.0128& 0.0159\\
~80 & 0.00778& 0.00975\\
~90& 0.00497 & 0.00630\\
100& 0.00332&0.00426\\
110& 0.00230& 0.00298\\
120&0.00164 & 0.00214\\
130&0.00120 & 0.00159\\
140&0.000901& 0.00120\\
150&0.000687&0.000922
\end{tabular}
\end{center}
\end{table}

\begin{figure}
\onefigure{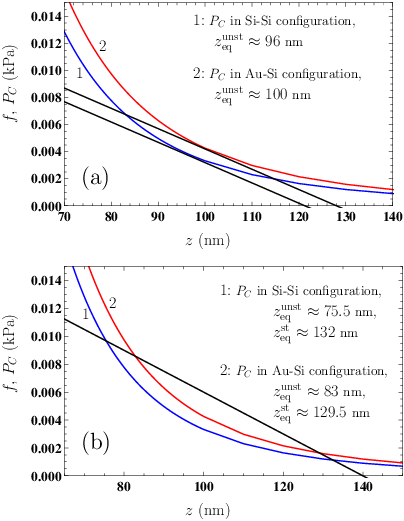}
\caption{Equilibrium positions of the sensor membrane made of Si above either
a Si substrate or an Au-coated Si substrate
with applied external pressures (a) $P=2.9818~$kPa and 2.9808~kPa, respectively,
 and (b) $P=2.973~$kPa.
See the text for further discussion.   }
\label{fg3}
\end{figure}

In fig.~\ref{fg3}(a,b), the Casimir pressures between the Si-Si and Si-Au
surfaces of a membrane and a substrate are shown as the functions of separation
by the lines 1 and 2, respectively. The bottom and top black lines in
fig.~\ref{fg3}(a) show the function $f(z)$ defined in Eq.~(\ref{eq4}) for the
spring constant $k=30~$N/m and for the values of applied pressure
$P=2.9818$ and 2.9808~kPa, respectively. For these applied pressures, there
is only one (unstable) solution of eq.~(\ref{eq4}) at
$z_{\rm eq}^{\rm unst}\approx 96~$nm for the sensor with an entirely Si
substrate and $z_{\rm eq}^{\rm unst}\approx 100~$nm for that with an Au-coated
substrate. For $P\geqslant 2.9818~$kPa and $P\geqslant 2.9808~$kPa the sensors
with Si-Si and Si-Au surfaces, respectively, collapse and become unusable.

The black line in fig.~\ref{fg3}(b) shows the function $f(z)$ plotted with
$P=2.979~$kPa. This value is smaller than both the limiting values indicated
above. As a result, there are both the stable and unstable equilibrium states
for the pressure sensors with Si and Au substrates. Thus, for sensors with
a Si substrate, one obtains $z_{\rm eq}^{\rm st}\approx 132~$nm and
$z_{\rm eq}^{\rm unst}\approx 75.5~$nm. For an Au-coated substrate,
$z_{\rm eq}^{\rm st}\approx 129.5~$nm and
$z_{\rm eq}^{\rm unst}\approx 83~$nm.

\section{Combined effect of Casimir and electric forces}

Here, the second configuration of pressure sensors is considered, where some
potential difference $V_0$ is applied between a membrane and a substrate.
In this case, eq.~(\ref{eq4}) is modified by the addition of the electric
pressure acting between a membrane and a substrate
\begin{equation}
f(z)\equiv\frac{kh}{S}-P-\frac{kz}{S}=P_C(z)+P_{\rm el}(z)=P_{\rm tot}(z),
\label{eq5}
\end{equation}
\noindent
where the magnitude of the electric pressure is
\begin{equation}
P_{\rm el}(z)=\frac{\epsilon_0V_0^2}{2z^2},
\label{eq6}
\end{equation}
\noindent
and
$\epsilon_0$ is the permittivity of vacuum.

\begin{figure}
\onefigure{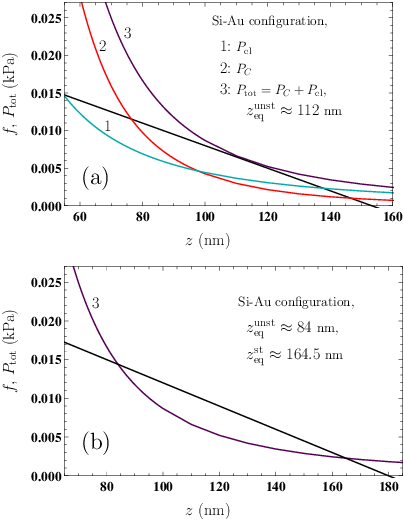}
\caption{Equilibrium positions of the sensor membrane made of Si above an
Au-coated Si substrate kept under the potential difference $V_0=0.1~$V
with applied external pressures (a) $P=2.977~$kPa and  (b) 2.973~kPa.
See the text for further discussion.  }
\label{fg4}
\end{figure}

We consider the solutions of this equation for $k=30~$N/m
and the applied voltage $V_0=0.1~$V assuming that the sensor substrate is
coated with an Au layer of $100~\muup$m thickness. In fig.~\ref{fg4}(a),
the electric, Casimir, and total pressure equal to their sum are shown as
the functions of separation by the lines 1, 2, and 3, respectively.
The function $f(z)$ defined in eq.~(\ref{eq5}) is shown by the straight
black line plotted for the external pressure $P=2.977~$kPa. As is seen
in fig.~\ref{fg4}(a), in this case eq.~(\ref{eq5}) has only one solution
$z_{\rm eq}^{\rm unst}\approx 112~$nm. Thus, if the external pressures
$P\geqslant 2.977~$kPa are applied to the sensor under consideration,
its membrane will stick to the substrate.

It is interesting to compare the relative roles of the electric and Casimir
forces at different separations between a sensor membrane and a
substrate. As is seen in fig.~\ref{fg4}(a), at separations exceeding 97 nm
the electric force shown by the line 1 is larger than the Casimir force (line 2).
However, at $a=97$~nm the values of both forces become equal and
at smaller separations the magnitude of the Casimir force significantly
exceeds the magnitude of the electric one.

In fig.~\ref{fg4}(b), the line labeled 3 again shows the sum of the
electric and Casimir forces, but the function $f(z)$ is plotted for
a slightly smaller applied pressure $P=2.973~$kPa.
From fig.~\ref{fg4}(b) it is seen that in this case the pressure
sensor has two equilibrium positions. One of them,
$z_{\rm eq}^{\rm unst}\approx 84~$nm is unstable, whereas another one,
$z_{\rm eq}^{\rm st}\approx 164.5~$nm, is stable. Therefore, the sensor
with the above parameters can be used for measuring the pressures below
2.977~kPa. With decreasing of the measured pressure, the equilibrium
value  $z_{\rm eq}^{\rm st}$ becomes larger which makes the role of the
Casimir force negligibly small.

\section{Conclusions and discussion}

In the foregoing, we presented several
simple examples illustrating the role played by the Casimir force in
MEMS and NEMS pressure sensors. It was shown that the Casimir force has a
significant impact on the functioning of pressure sensors when, under the
action of external pressure, the separation distance between a membrane
and a substrate drops to about 100 nm or even less. This happens under a
certain relationship between the spring constant, initial height of the
membrane above a substrate, and the external pressure.

According to our results, at some maximum value of the measured external
pressure, the sensor finds itself in the position of unstable equilibrium.
At this and larger measured pressures, the sensor membrane collapses onto
the substrate. As a result, the sensor becomes unsuitable for further use.
At all smaller pressure values, there are two equilibrium positions of the
sensor membrane under an impact of the Casimir force, one of which is
unstable and another one is stable, at shorter and larger heights above
a substrate, respectively. The stable equilibrium position can be used for
pressure measurements without risk of damaging the sensor.

{In this Letter, all calculations have been made for the case of
smooth Au and Si surfaces. However, as was noted in Introduction, all
surfaces of the elements of microdevices are characterized by some roughness.
This roughness makes an impact on both the Casimir and electric forces and
should be taken into account in their computations. In fact, the surfaces
of micro- and, especially, nanodevices should be made sufficiently smooth.
As an example, the stochastic roughness of the Au-coated surfaces used in
precise measurements of the Casimir force by means of an atomic force
microscope is characterized by the root-mean-square amplitude $\delta$ equal
to approximately 1.1~nm \cite{14.11,14.12,14.13}. Several theoretical methods
were developed on how to account for the role of surface roughness in force
calculations (see, e.g., \cite{22,23,24,26,rg1,rg2,rg3,rg4,rg5}).
When constructing a nanodevice, the roughness profiles should be investigated
by means of an atomic force microscope and its impact on the forces found using
these methods. In any case, under a condition that $\delta$ is much smaller than
the characteristic separation between the elements of a nanodevice, the account
of roughness does not change the qualitative results presented above, but only
slightly changes the positions of stable and unstable equilibrium.
}

{The results of this Letter}
are important for the design of new generations of MEMS and
NEMS pressure sensors with further decreased dimensions. The developed approach
makes it possible to find the values of all relevant parameters in such
a way in order to avoid the damage or destruction of sensor under the
influence of the Casimir force.

\acknowledgments
The work of G.L.K. and V.M.M. was partially funded by the
Ministry of Science and Higher Education of Russian Federation
as part of the World-Class Research Center program: Advanced Digital Technologies
(contract No. 075-15-2022-311 dated April 20, 2022).
A.S.K. and V.V.L. thank the State Assignment for Basic Research
(project FSEG-2023-0016) for financial support.
The work of
V.M.M.~was also partially carried out in accordance with the Strategic
Academic Leadership Program "Priority 2030" of the Kazan Federal
University.


\begin{thebibliography}{99}
\bibitem{1}
Elwenspoek M. and  Wiegerink R.,
{\it  Mechanical Microsensors}
(Springer, Berlin) 2001.
\bibitem{2}
Gardner J.~W., Varadan V.\ K. and Awadelkarim O.\ O.,
{\it Microsensors, MEMS and Smart Devices}
(Wiley, Chishester) 2013.
\bibitem{2.1}
Zhao X., Tsai J. M., Cai H., Ji X. M., Zhou J., Bao  M. H., Huang Y. P., Kwong D. L. and Liu A. Q.,
{\it Optics Express}, {\bf 20} (2012) 8535.
\bibitem{2.2}
Lee D., Kim J., Kim H., Heo H., Park, K. and Lee Y.,
{\it Nanoscale}, {\bf 10} (2018) 18812.
\bibitem{2.3}
Gao Y., Lu C., Guohui Y., Sha J., Tan J. and Xuan F.,
{\it Nanotechnology}, {\bf 30} (2019) 325502.
\bibitem{2.4}
Chang Y., Zuo J., Zhang H. and Duan X.,
{\it Nanotech. Prec. Engin.}, {\bf 3} (2020) 43.
\bibitem{2.5}
Mahata C., Algadi H., Lee J., Kim S. and Lee T.,
{\it Measurement}, {\bf 151} (2020) 107095.
\bibitem{2.6}
Mohammed M. K., Al-Nafiey A. and Al-Dahash, G.,
{\it Nano Biomed. Eng.}, {\bf 13} (2021) 27.
\bibitem{2.7}
Tang R., Lu F., Liu L., Yan Y., Du Q., Zhang B., Zhou T. and Fu H.,
{\it Nano}, {\bf 2} (2021) 1874.
\bibitem{2.8}
Romijn J., Dolleman R. J., Singh M., van der Zant H. S. J., Steeneken P. J., Sarro P. M. and Vollebregt S.,
{\it Nanotechnology}, {\bf 32} (2021) 335501.
\bibitem{2.9}
Shin Y.-K., Shin Y., Lee J. W. and Seo M.-H.,
{\it Biosensors}, {\bf 12} (2022) 952.
\bibitem{2.10}
Bao Y., Xu J., Guo R. and Ma J.
{\it Progr. Chem.}, {\bf 35} (2023) 709.
\bibitem{3}
Gruca G., Chavan D., Rector J., Heeck K. and Iannuzzi D.,
{\it Sens. Actuators A: Phys.}, {\bf 190} (2013)  77.
\bibitem{4}
Mishra R.~B. and Kumar S.~S.,
{\it J. Phys.: Conf. Ser.}, {\bf 1240}  (2019) 012068.
\bibitem{5}
Pisco M. and  Cusano A.,
{\it Sensors}, {\bf 20} (2020) 4705.
\bibitem{6}
Han X., Huang M., Wu Z. et al.,
{\it Microsystems \& Nanoengineering}, {\bf 9} (2023) 156.
\bibitem{7}
Chen S.,  Qin J., Lu Y., Xie B., Wang J.,
Chen D. and Chen J.,
{\it Micromachines}, {\bf 14} (2023) 441.
\bibitem{8}
Qian P., Yu Z., Yu J.,  Lu Y.,  Xie B., Chen J., Chen D.
and Wang J.,
{\it Microsystems \& Nanoengineering}, {\bf 10}  (2024) 38.
\bibitem{9}
Tutaev A.~T., Koslov A.~S., Belyaev J.~V., Loboda V.~V., Bellavin M.~A.
and Korotkov A.~S.
{\it IEEE Sensors J.}, {\bf 24} (2024) 7395.
\bibitem{10}
Casimir, H.~B.~G.,
{\it Proc. Kon. Ned. Akad. Wet. B}, {\bf 51} (1948) 793.
\bibitem{11}
Buks E. and Roukes M.~L.,
{\it Phys. Rev. B}, {\bf 63} (2001) 033402.
\bibitem{12}
Buks E. and Roukes M.~L.,
{\it Europhys. Lett.}, {\bf 54} (2001) 220.
\bibitem{12a}
Yang R., Qian J. and Feng P.~X.-L.,
{\it Small}, {\bf 16} (2020) 2005594.
\bibitem{12b}
Schmidt F., Callegari A., Daddi-Moussa-Ider A. et al.,
{\it Nature Phys.}, {\bf 19} (2023) 271.
\bibitem{13}
Chan H.~B., Aksyuk V.~A., Kleiman R.{\ }N., Bishop D.{\ }J.
and Capasso F.,
{\it Science}, {\bf 291} (2001) 1941.
\bibitem{14}
Chan H.~B., Aksyuk V.~A., Kleiman R.{\ }N., Bishop D.{\ }J.
and Capasso F.,
{\it Phys. Rev. Lett.}, {\bf 87} (2001) 211801.
\bibitem{14.1}
Decca R.~S., Fischbach E., Klimchitskaya G.~L., Krause D.~E.,
L\'{o}pez D. and Mostepanenko V.~M.,
{\it Phys. Rev. D}, {\bf 68} (2003) 116003.
\bibitem{14.2}
Decca R.~S., L\'{o}pez D., Fischbach E., Klimchitskaya G.~L., Krause D.~E.
and Mostepanenko V.~M.,
{\it Ann. Phys. (NY)}, {\bf 318} (2005) 37.
\bibitem{14.3}
Decca R.~S., L\'{o}pez D., Fischbach E., Klimchitskaya G.~L., Krause D.~E.
and Mostepanenko V.~M.,
{\it Phys. Rev. D}, {\bf 75} (2007) 077101.
\bibitem{14.4}
Decca R.~S., L\'{o}pez D., Fischbach E., Klimchitskaya G.~L., Krause D.~E.
and Mostepanenko V.~M.,
{\it Eur. Phys. J. C}, {\bf 51} (2007) 963.
\bibitem{14.5}
Chang C.-C., Banishev A.~A., Castillo-Garza R.,
Klimchitskaya G.~L., Mostepanenko V.\ M. and Mohideen U.,
{\it Phys. Rev. B}, {\bf 85} (2012) 165443.
\bibitem{14.6}
Banishev A.~A., Chang C.-C.,
Klimchitskaya G.~L., Mostepanenko V.\ M. and Mohideen U.,
{\it Phys. Rev. B}, {\bf 85} (2012) 195422.
\bibitem{14.7}
Banishev A.~A.,
Klimchitskaya G.~L., Mostepanenko V.\ M. and Mohideen U.,
{\it Phys. Rev. Lett.}, {\bf 110} (2013) 137401.
\bibitem{14.8}
Banishev A.~A.,
Klimchitskaya G.~L., Mostepanenko V.\ M. and Mohideen U.,
{\it Phys. Rev. B}, {\bf 88} (2013) 155410.
\bibitem{14.9}
Bimonte G., L\'{o}pez D. and Decca R.\ S.,
{\it Phys. Rev. B}, {\bf 93} (2016) 184434.
\bibitem{14.10}
Bimonte G., Spreng B., Maia Neto P. A., Ingold G.-L., Klimchitskaya G.~L.,
Mostepanenko V.\ M. and Decca R.~S.,
{\it Universe}, {\bf 7} (2021) 93.
\bibitem{14.11}
Xu J.,
Klimchitskaya G.~L., Mostepanenko V.\ M. and Mohideen U.,
{\it Phys. Rev. A}, {\bf 97} (2018) 032501.
\bibitem{14.12}
Liu M., Xu J.,
Klimchitskaya G.~L., Mostepanenko V.\ M. and Mohideen U.,
{\it Phys. Rev. B}, {\bf 100} (2019) 081406(R).
\bibitem{14.13}
Liu M., Xu J.,
Klimchitskaya G.~L., Mostepanenko V.\ M. and Mohideen U.,
{\it Phys. Rev. A}, {\bf 100} (2019) 052511.
\bibitem{14.14}
Chen F., Mohideen U., Klimchitskaya G.~L. and Mostepanenko V.~M.,
{\it Phys. Rev. A}, {\bf 74} (2006) 022103.
\bibitem{14.15}
Chen F., Klimchitskaya G.~L., Mostepanenko V.~M. and Mohideen U.,
{\it Phys. Rev. Lett.}, {\bf 97} (2006) 170402.
\bibitem{14.16}
Chen F., Klimchitskaya G.~L., Mostepanenko V.~M. and Mohideen U.,
{\it Opt. Express}, {\bf 15} (2007)  4823.
\bibitem{14.17}
Chen F., Klimchitskaya G.~L., Mostepanenko V.~M. and Mohideen U.,
{\it Phys. Rev. B}, {\bf 76} (2007) 035338.
\bibitem{14.18}
Obrecht J.~M., Wild R.~J., Antezza M., Pitaevskii L.~P.,
Stringari S. and Cornell E.~A.
{\it Phys. Rev. Lett.}, {\bf 98} (2007) 063201.
\bibitem{14.19}
Chang C.-C., Banishev A.~A., Klimchitskaya G.~L., Mostepanenko V.~M. and Mohideen U.,
{\it Phys. Rev. Lett.}, {\bf 107} (2011) 090403.
\bibitem{14.20}
Banishev A.~A., Chang C.-C., Castillo-Garza R., Klimchitskaya G.~L., Mostepanenko V.~M.
and Mohideen U.,
{\it Phys. Rev. B}, {\bf 85} (2012) 045436.
\bibitem{15}
Klimchitskaya G.~L., Mohideen U. and Mostepanenko V.\ M.,
{\it Rev. Mod. Phys.}, {\bf 81} (2009) 1827.
\bibitem{16}
Klimchitskaya G.~L., Mohideen U. and Mostepanenko V.\ M.,
{\it Int. J. Mod. Phys. B}, {\bf 25} (2011) 171.
\bibitem{17}
Bordag M., Klimchitskaya G.~L., Mohideen U. and
Mostepanenko V.\ M.,
{\it Advances in the Casimir Effect}
(Oxford University Press, Oxford) 2015.
\bibitem{18}
Woods L.~M., Dalvit D.~A.~R., Tkatchenko A., Rodriguez-Lopez P.,
Rodriguez A.\ W. and Podgornik R.,
{\it Rev. Mod. Phys.}, {\bf 88} (2016) 045003.
\bibitem{19}
Mostepanenko V.~M.,
{\it Universe}, {\bf 7} (2021) 84.
\bibitem{20}
Barcenas J., Reyes L. and Esquivel-Sirvent R.,
{\it Appl. Phys. Lett.}, {\bf 87} (2005) 263106.
\bibitem{21}
Esquivel-Sirvent R. and P\'{e}rez-Pascual R.,
{\it Eur. Phys. J. B}, {\bf 86} (2013) 467.
\bibitem {22}
Palasantzas G.,
{\it J. Appl. Phys.}, {\bf 101} (2007) 053512.
\bibitem {23}
Palasantzas G.,
{\it J. Appl. Phys.} {\bf 101} (2007) 063548.
\bibitem{24}
Broer W., Palasantzas G., Knoester G. and Svetovoy V.\ B.,
{\it Phys. Rev. B}, {\bf 87}  (2013) 125413.
\bibitem{25}
Sedighi M., Broer W., Palasantzas G. and Kooi B.\ J.,
{\it Phys. Rev. B}, {\bf 88} (2013) 165423.
\bibitem{26}
Broer W., Waalkens H., Svetovoy V.\ B., Knoester J.
and Palasantzas G.,
{\it Phys. Rev. Appl.}, {\bf 4} (2013) 054016.
\bibitem{27}
Zou J., Marcet Z., Rodriguez  A.\ W., Reid M.\ T.\ H.,
McCauley A.\ P., Kravchenko I.\ I.,  Lu T., Bao Y.,
Johnson S.\ G. and  Chan H.\ B.,
{\it Nat. Commun.}, {\bf 4} (2013) 1845.
\bibitem{28}
Tang L., Wang M., Ng C.~Y., Nicolic M., Chan C.\ T.,
 Rodriguez A.{\ }W.  and Chan H.\ B.,
{\it Nature Photonics}, {\bf 11} (2017) 97.
\bibitem{29}
Inui N.,
{\it J. Appl. Phys.}, {\bf 122}  (2017) 104501.
\bibitem{30}
Klimchitskaya G.~L., Mostepanenko V.\ M., Petrov V.\ M. and Tschudi T.,
{\it Phys. Rev. Appl.}, {\bf 10} (2018) 014010.
\bibitem{31}
Stange A., Imboden M., Javor J., Barrett L.\ K. and  Bishop D.\ J.,
{\it Microsystems \& Nanoengineering}, {\bf 5}  (2019) 14.
\bibitem{32}
Javor J., Yao Z., Imboden M.,  Campbell D.\ K. and  Bishop D.\ J.,
{\it Microsystems \& Nanoengineering}, {\bf 7}  (2021) 73.
\bibitem{33}
Sircar A.,  Patra P.\ K. and  Batra R.\ C.,
{\it Proc. R. Soc. A}, {\bf 476} (2020) 20200311.
\bibitem{34}
Lifshitz E.~M.,
{\it Zh. Eksp. Teor. Fiz.}, {\bf 29} (1955) 94
[{\it Sov. Phys. JETP}, {\bf 2} (1956) 73].
\bibitem{34a}
Dzyaloshinskii I.~E., Lifshitz E.~M. and Pitaevskii L.\ P.,
{\it Usp. Fiz. Nauk}, {\bf 73} (1961) 381
[{\it Adv. Phys.}, {\bf 10} (1961) 165].
\bibitem{34b}
Lifshitz E.~M. and Pitaevskii L.\ P.,
{\it Statistical Physics, Part II}
(Pergamon, Oxford) 1980.
\bibitem{35}
Palik, E.D. (Ed.)
{\it Handbook of Optical Constants of Solids}
(Academic Press, New York) 1985.
\bibitem{36}
Klimchitskaya G.~L. and Mostepanenko V.~M.,
{\it Physics}, {\bf 5} (2023) 952.
\bibitem{37}
Klimchitskaya G.~L., Mostepanenko V.~M. and Svetovoy V.~B.,
{\it Europhys. Lett.}, {\bf 139} (2022) 66001.
\bibitem{38}
Klimchitskaya G.~L., Mostepanenko V.~M. and Svetovoy V.~B.,
{\it Universe}, {\bf 8} (2022) 574.
\bibitem{liq1}
Nie Z., Kwak J.~W., Han M. and Rogers J.~A.
{\it Advanced Materials}, {\bf 35} (2023) 2205609.
\bibitem{liq2}
Meena K.~V. and Sankar A.~R.
{\it IEEE Sensors J.}, {\bf 21} (2021) 10241.
\bibitem{gas1}
Li C., Cordovilla F. and Oca\~{n}a J.~L.
{\it Solid-State Electronics}, {\bf 139} (2018) 39.
\bibitem{gas2}
Loboda V.~V. and Salamatova U.~V.
{\it Computing, Telecommunications and Control}, {\bf 14} (2021) 65.
{
\bibitem {rg1}
Bordag M., Klimchitskaya G.L. and Mostepanenko V.M.
{\it Int. J. Mod. Phys. A}, {\bf 10 } (1995) 2661.
\bibitem {rg2}
Maia Neto P.A., Lambrecht A. and Reynaud S.
{\it Phys. Rev. A}, {\bf 72} (2005) 012115.
\bibitem {rg3}
Broer W., Palasantzas G., Knoester J. and Svetovoy V.B.
{\it Erophys. Lett.}, {\bf 95} (2011) 30001.
\bibitem {rg4}
Svetovoy V.B. and Palasantzas G.
{\it Adv. Coll. Interface Sci.}, {\bf 216} (2015) 1.
\bibitem {rg5}
Makeev M.A.
{\it Surf. Sci.}, {\bf 663} (2017) 88.
}

\end{thebibliography}
\end{document}